\documentclass[pre,amsmath,amssymb,twocolumn]{revtex4}

\usepackage{graphicx}
\usepackage{dcolumn}
\usepackage{bm}

\usepackage{epsfig}

\begin{document}

\title{Fluid-bicontinuous gels stabilized by interfacial colloids:\\ low and high molecular weight fluids}
\author{P~S~Clegg}
\email{pclegg@ph.ed.ac.uk}
\affiliation{SUPA, School of Physics, University of Edinburgh, Mayfield Road, Edinburgh, EH9 3JZ, UK}

\begin{abstract}
Carefully tuned composite materials can have properties wholly unlike their separate constituents. We review the development of one example: colloid-stabilized emulsions with bicontinuous liquid domains. These non-equilibrium structures resemble the sponge mesophase of surfactants; however, in the colloid-stabilized case the interface separating the liquid domains is itself semi-solid. The arrangement of domains is created by arresting liquid-liquid phase separation via spinodal decomposition. Dispersed colloids exhibiting partial wettability become trapped on the newly created interface and jam together as the domains coarsen. Similar structures have been created in polymer blends stabilized using either interfacial nanoparticles or clay platelets. Here it has been possible to create the domain arrangement either by phase separation or by direct mixing of the melt. The low molecular-weight liquid and polymer based structures have been developed independently and much can be learnt by comparing the two.
\end{abstract}

\maketitle

\section{Introduction}

Soft-composite materials can exhibit dramatic static and dynamic properties~\cite{Torquato02}~\cite{Larson99}, although choosing components and processing techniques, so as to create desired characteristics, is complex. Historically, composites have been developed by trial and error: an approach that is far from being efficient. Currently, in some specific areas, rational generic approaches to creating novel soft materials are beginning to be identified, such as designing novel subunits~\cite{vanBlaaderen06}~\cite{Glotzer07} or making use of the chemistry of organized matter~\cite{Colfen03}. Another generic approach is to create materials that are arrested out of equilibrium leading to robust soft composites with tunable properties~\cite{Stratford05}. We focus on a class of arrested three-component soft solids with an architecture that leads to novel properties and is likely to lead to numerous applications. Our route is to create arrested composites via the stabilization of convoluted fluid-fluid interfaces~\cite{Stratford05} and it can be applied to binary fluids and polymer blends both containing dispersed particles. For binary fluids the resulting properties are wholly different to those of the initial ingredients~\cite{Herzig07}. For polymer blends, by contrast, it is typically a stable domain arrangement, which combines the characteristics of the constituents, that is required~\cite{Paul80}.

\begin{figure}
\centerline{\includegraphics[scale=0.8]{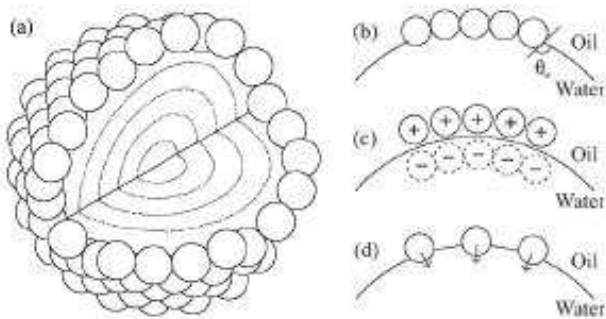}}
\caption{\label{Minimum} (a) Cut-away of a colloid-stabilized droplet: the inside and outside surfaces of the colloids are identical (after Israelachvili). The arrangement of colloids can be controlled by: (b) the wetting angle, $\theta_w$, (c) image charges in the high-dielectric constant medium and (d) dipole formation (arrows) due to surface charges and an asymmetric distribution of counterions.}
\end{figure}

Recently there has been a resurgence of interest in the behaviour of homogeneous colloids at fluid-fluid interfaces~\cite{Binks06}. Colloids that exhibit partial wettability with a pair of immiscible fluids can become trapped at interfaces between them and can be used as emulsifiers (Fig.~\ref{Minimum}a). The trapping is due to the reduction of the shared area (and hence interfacial energy) between the fluids. The energy barrier holding the particles to the interface is large and consequently, the properties of emulsions stabilized by interfacial colloids are quite different to those of emulsions stabilized by amphiphilic molecules. One notable example is that colloid-stabilized emulsions will invert in response to a change in the volume fraction of fluids alone~\cite{Binks00c}.

Colloid-stabilized interfaces are novel soft materials in their own right~\cite{Aveyard00}~\cite{Stancik02}. A low area fraction of colloids will form aggregates or a tenuous gel depending on the direct colloid-colloid interactions and on the additional interactions that arise due to the interface~\cite{Bergstrom06}. At high area fractions the interface will eventually either jam or crystallize. On a spherical surface, as for emulsion droplets, crystalline packing will necessarily involve defects~\cite{Bausch03}. An essential feature, for studies of non-spherical droplets, is that a surface covered with a jammed or crystalline layer will have solid character on length scales larger than the colloids and will be liquid in the interstices between the colloids. The solidity implies that the surface can support macroscale variations in mean curvature, which has been demonstrated experimentally~\cite{Subramaniam05}.

The low molecular-weight liquid studies reviewed here use liquid-liquid demixing via spinodal decomposition to create a bicontinuous arrangement of domains. This route to creating novel particle-stabilized emulsions was first explored via large-scale computer simulations~\cite{Stratford05}. Colloids are swept up on the freshly created interfaces and, as the domains coarsen, the area fraction of colloids increases and the colloids jam together. These materials are known as bicontinuous interfacially jammed emulsion gels (bijels). Clegg and coworkers first fabricated thin slabs experimentally using alcohol-oil mixtures~\cite{Clegg07} and then Herzig and coworkers went on to create fully three-dimensional structures using water-lutidine mixtures~\cite{Herzig07}. In both cases silica colloids were used to give long-term stability to the interface.

By spin-casting a pair of polymers, initially dispersed in a common solvent, Chung and coworkers have been able to arrest the spinodal pattern in a thin film using neutrally wetting nanoparticles~\cite{Chung05}. There are comparatively few partially miscible pairs of polymers, but fortunately, it is easier to create bicontinuous domain patterns via phase inversion or melt mixing of polymer blends~\cite{Potschke03}. Phase inversion occurs in a polymer blend under shear if the polymer properties depend on temperature, shear rate, etc, while melt mixing can lead to a bicontinuous arrangement of domains via repeated stretching and folding followed by rupture. Recent research has shown that organically modified clay will sequester to the interface of blends~\cite{Fang07}. These clay platelets have subsequently been used by Si and coworkers to arrest three-dimensional bicontinuous domain arrangements created via the melt mixing of polymer blends~\cite{Si06}.

This review attempts to combine two fields that have previously been immiscible; the lack of common references that both cite presents a challenge to literature searches and may mean that some work has been inadvertently overlooked. This review follows after those of references~\cite{Boker07, Binks06, Zeng06, Miller06, Wang05, Aveyard03, Binks02, Tambe94, Menon88}  concerning aspects of colloid-stabilized interfaces and emulsions and references including~\cite{Balazs06, Potschke03, Lipatov02, Utracki98, DiLorenzo97, Lyngaaejorgensen91, Barlow81, Paul80} on polymer blends; in both fields we focus on recent results where bicontinuous arrangements of domains are stabilized. In section II we discuss the mechanisms by which colloids become trapped at the fluid-fluid interfaces, standard routes to forming droplet emulsions and emulsion inversion. In section III we consider interfacial elasticity and the occurrence of non-spherical emulsion droplets. In section IV we detail recent simulations and experiments concerning the arrest of liquid-liquid phase separation using colloidal particles. Analogous studies using polymer blends are described in section V. In section VI we look at future directions, including what each field may have to teach the other.

\section{Colloid-stabilized emulsions}

The energetics of the trapping of colloids at fluid-fluid interfaces is relatively straight-forward. By contrast, current research on the interactions between colloids on interfaces, the kinetics of trapping and the inversion of colloid-stabilized emulsions, shows these are far more challenging to understand. These topics underpin the subsequent parts of this review.

\subsection{Colloids and fluid-fluid interfaces} 

If a colloidal particle has intermediate wettability with two fluids then it can sequester to an interface between them (Fig.~\ref{Minimum}b), reducing the shared area and hence the energy cost~\cite{Binks06}. The wettability is described by a three-phase contact angle, $\theta_w$, and the change in free energy when the spherical colloid moves from the interface into the preferred fluid is
\begin{equation}
\Delta G_{int} = \pi r^2 \gamma_{ow} (1 - |\cos{\theta_w}|)^2,
\label{Energy}
\end{equation}

\noindent where $r$ is the radius of the colloid and $\gamma_{ow}$ is the interfacial tension between the two liquids. The energy well is deepest for neutral wetting ($\theta_w = 90^{\circ}$), large particles and a high interfacial tension. A second mechanism, the electrostatic trapping of particles close to an interface, has also been considered. Earnshaw calculated the depth of the energy well for a charged particle in an aqueous phase due to the distortion of the counterion cloud as an interface is approached~\cite{Earnshaw86}. More recently Leunissen and coworkers~\cite{Leunissen07} experimented with charged particles in an oil phase: if poly(methyl methacrylate) (PMMA) particles are dispersed in droplets of an oil phase (cyclohexyl bromide and decalin mixture) where the continuous phase is water, the particles sequester to the interface in spite of the fact that the wetting angle $\theta_w \sim 180^{\circ}$. This is because the water mops up ions from the oily solvent leaving a +450e charge on each colloid. These particles are then simultaneously strongly attracted to their image charge (Fig.~\ref{Minimum}c) and repelled by the interface which is thought to have the same charge. Because the charge of the colloid and its image effectively cancel, a close-packed coating can form around a droplet. Subsequent computer simulations suggest that the standard interfacial trapping (equation (1)) may also play a role in this system~\cite{Zwanikken07}. 

Populations of particles trapped by interfacial tension (equation (1)) exhibit a broad range of behaviour including long-range order~\cite{Horozov03}~\cite{Horozov05}~\cite{Park07}, aggregation in fractal clusters~\cite{Robinson92} and mesostructure formation~\cite{Ruiz-Garcia98}. These are evidence of the diverse interactions that can play a role depending on the details of the two fluids and the colloids~\cite{Kralchevsky00}~\cite{Kralchevsky01}~\cite{Fernandez-Toledano06}. Ionic surface groups will tend to dissociate in a polar solvent leading to exposed charges and associated interactions. The asymmetric arrangement of charges and counterions close to an interface can result in a charge dipole~\cite{Pieranski80}~\cite{Hurd85} making the two-dimensional case quite different to that in three dimensions (Fig.~\ref{Minimum}d). By means that are not yet fully understood exposed charges can appear in the non-polar phase~\cite{Aveyard00} leading to electrostatic interactions through the low dielectric constant medium. Van der Waals attractions are more complicated for particles at an interface due to the two different media. A further interaction, mediated by the fluid-fluid interface, arises for particles with a rough surface or otherwise inhomogeneous wetting properties~\cite{Stamou00}~\cite{Kralchevsky01b}. This effect can be large compared to thermal energies~\cite{Arditty05}. Two rough particles will attempt to orient themselves such that the distortion to the interface is minimized~\cite{Stamou00}. Some of the behaviour of colloids at fluid-fluid interfaces has yet to be explained.

\subsection{Mixing liquids and colloid-stabilized emulsions} 

A layer of particles trapped via the mechanism described by equation~(\ref{Energy}) is able to stabilize droplets of an immiscible dispersed phase. Stabilization can occur via a number of routes: firstly, if a droplet is completely covered by particles that are jammed together, then this acts as a mechanical barrier against coalescence~\cite{Tambe94}~\cite{Midmore98}~\cite{Binks02} (unlike with amphiphilic surfactants, a densely packed interfacial layer of particles has solid-like character). Secondly, if the colloids have sufficiently long-range repulsions an open crystal structure can form over the surface of the droplet, and when two droplets approach they share the layer of colloids between them (bridging)~\cite{Stancik04}~\cite{Horozov05}~\cite{Horozov06}. This has been demonstrated both for droplet-flat interface collisions and for droplet-droplet collisions. The particles that bridge the interfaces tend to aggregate to form a densely-packed disk. The bridging geometry eliminates the asymmetry in the particle charge distribution and reduces the shared oil-particle surface area~\cite{Stancik04} which will both reduce the repulsion between particles. Recent experiments suggest that aggregation of the bridging particles in the contact region is largely due to the lateral capillary attraction~\cite{Xu07}. Thirdly, droplets can also be stabilized by a partial coating of  particles~\cite{Vignati03}~\cite{Tarimala04} which tend to move to prevent exposed regions of droplets coming into contact; this may be another example of bridging by shared particles. These mechanisms are all quite different to those for amphiphilic surfactants; however, the colloid wetting angle plays a similar role to the hydrophilic-lipophilic balance: the wetting angle at the liquid-liquid-colloid contact line will influence the curvature of the liquid-liquid interface because the bulk of the colloid will sit in the liquid where its surface energy is lowest~\cite{Binks02}.

We are interested in colloids being swept up by newly formed liquid-liquid interfaces and in the subsequent evolution of these interfaces. Sweeping up is complex and has been explored in the context of froth flotation~\cite{Nguyen06}~\cite{Nguyen06b}. In spite of the deep energy well for a neutrally-wetting colloid at an interface (equation (1)) there can be a barrier caused by line tension and delays to attachment associated with film thinning and the kinetics of forming a three-phase contact line. The interfaces evolve as partially-coated droplets coarsen by coalescence which (except in unusual circumstances) stops when the droplets have a complete coating of colloids. The coalescence rate decreases with droplet size as arrest is approached~\cite{Arditty03}, and this size-dependent coalescence rate leads to a droplet size distribution that is surprisingly uniform.

\subsection{Inversion of colloid-stabilized emulsions} 

\begin{figure}
\centerline{\includegraphics[scale=0.8]{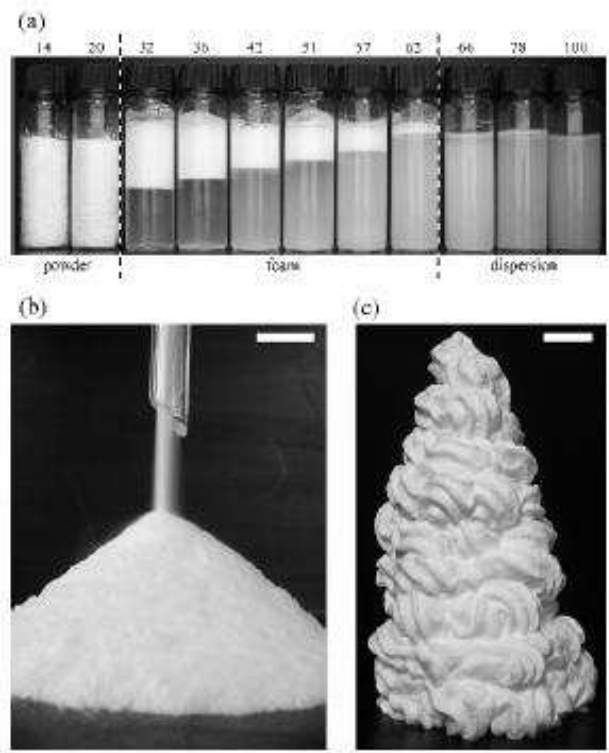}}
\caption{\label{Inversion} Transitional inversion: (a) sequence of air-water samples mixed with 2\% weight silica particles that become progressively more hydrophilic to the right. The numbers above are the (\%) SiOH content on the particle surface; (b) a free-flowing powder of water stabilized with hydrophobic particles (20\% SiOH); (c) water-air foam stabilized with slightly more hydrophilic particles (32\% SiOH). The scale bars are 1~cm. Reprinted by permission from Macmillan Publishers Ltd: Nature Materials~\cite{Binks06b}, copyright 2006.}
\end{figure}
Inversion is of interest here because bicontinuity may be associated with the composition at which an emulsion inverts. Colloid-stabilized emulsions invert in response to changing the surface chemistry of the colloid, transitional inversion~\cite{Binks00t}, or varying the relative volumes of the dispersed and continuous phases, catastrophic inversion~\cite{Binks00c}.  Transitional inversion has been achieved in two distinct ways~\cite{Aveyard03} with markedly different results: firstly, the wetting angle of all of the particles can be changed together (Fig.~\ref{Inversion}); secondly, the average wetting angle can be tuned by varying the proportions of two types of particles with very different wetting properties. For a particle population with a uniform wetting character the approach to inversion is characterized by the extreme stability of the emulsions and the droplet size passing through a sharp minimum. This is surprising because the curvature induced by the particles should be zero at the inversion point. For a bimodal distribution of wetting angles the droplet size does increase towards the inversion point though the reason for the difference in droplet size remains unclear. Catastrophic inversion contradicts the idea that the fluid that preferentially wets the surface of the colloid always forms the continuous phase~\cite{Binks02}. If the volume fraction of the continuous phase is decreased sufficiently the emulsion will invert regardless of the wetting angle~\cite{Binks00c}. (A similar switch in the sign of the mean curvature is seen for a particle-coated pendant drop being retracted using a syringe, see Fig.~9 of ref.~\cite{Asekomhe05}.) There is a discontinuous change in droplet size across the inversion point, and the viscosity of a droplet emulsion is also observed to increase approaching this composition, although the peak viscosity occurs before the inversion is actually reached. 

Emulsion inversion remains an area of active research both experimentally and theoretically. It has been possible to synthesise colloids with temperature dependent surface chemistry such that emulsions made at different temperatures span the range through inversion and at the inversion point a multiple emulsion is seen~\cite{Binks05}. Further, Binks and Murakami~\cite{Binks06b} have shown that water-air mixtures stabilized by interfacial silica can also be made to exhibit both catastrophic and transitional inversion (Fig.~\ref{Inversion}). Kralchevsky and coworkers~\cite{Kralchevsky05} have considered theoretically the combined effects of emulsion composition, droplet curvature and the wetting angle. Using a thermodynamic model they find that the liquid that preferentially wets the particle will form the continuous phase provided that it is in excess. When the proportion of this liquid is reduced the emulsion undergoes catastrophic inversion to droplets that will tend to floculate. The model predicts that catastrophic inversion will always occur at equal volumes of the two liquids which contradicts some experimental results.

\section{Interfacial elasticity and non-spherical emulsion droplets}

As part of the effort to glean how emulsions are stabilized by colloids, various groups have studied the mechanical properties of flat particle-laden interfaces and emulsion droplets. This research has primarily focused on the interfaces between low molecular-weight fluids and forms the background to our fabrication of bijels.

\subsection{Properties of flat colloid-coated interfaces} 

Colloids have a quite different effect on interfacial tension than soluble amphiphiles~\cite{Clegg07}. Since the energy barrier for colloids (equation (1)) is very large and the objects themselves are mesoscopic, small changes in the area of the interface are not accompanied by a change in the number of colloids attached. Consequently, an increase in the area of the interface will lead to increased contact area between bare fluids and an associated energy cost. A reduction in area requires the expulsion of colloids also giving an energy increase. This situation can be described by an interfacial elasticity model rather than by the interfacial tension description suitable for soluble amphiphiles. Experiments are required to discover the detailed mechanics of the interface.
\begin{figure}[b]
\centerline{\includegraphics[scale=0.65]{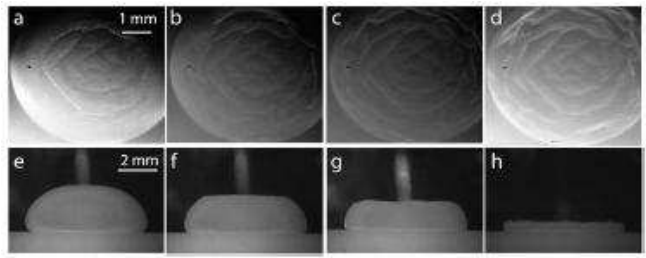}}
\caption{\label{Crumpling} A sequence of images (top view (a-d) and side view (e-h)) as a  water droplet surrounded by decane is deflated; the droplet interface is stabilized by a layer of polystyrene colloids and the onset of buckling is clearly visible. Reprinted with permission from~\cite{Xu05}. Copyright 2005 American Chemical Society.}
\end{figure}

Rheology has been used to demonstrate that an interface packed with colloids is a viscoelastic medium. Indirect characterization of the mechanical properties was carried out in a Langmuir trough~\cite{Aveyard00b} showing that a crystalline monolayer of colloids will crumple as the surface pressure is raised. Crumpling maintains the area per particle within the monolayer; the monolayer is folded to accommodate the reduced area of the trough. The periodicity of the buckling pattern can be used to infer a value of the Young's modulus ($E \sim \gamma_{ow} / r$) that agrees well with a simple model~\cite{Vella04}~\cite{Poisson}. For non-close packed charged colloids the packing first distorts prior to buckling~\cite{Aveyard00}. Studies have also probed the response of a colloid-laden interface to a variety of flow regimes~\cite{Stancik02}~\cite{Stancik03}. At low shear rates the colloidal planes tend to slide past one another (hydrodynamic forces win over colloid-colloid interactions) while at high shear rates domains of crystalline colloidal packing rotate together. Studying such particle-laden layers in the context of an emulsion requires a consideration of the role of curvature.

\subsection{Character of the interfaces of colloid-coated droplets} 

On the colloidal scale, packing defects necessarily accompany crystallization on the surface of a spherical droplet~\cite{Bausch03}~\cite{Lipowsky05}. The situation becomes more complicated for larger droplets covered by small particles where the defects are masked by grain boundaries which reduce the elastic energy penalty. Disordered packing and glassy behaviour may become the norm for other interfaces especially those that exhibit constant positive or negative Gaussian curvature~\cite{Nelson02}~\cite{Modes07}. On the macroscale, the rheological and osmotic properties of high volume-fraction colloid-stabilized emulsions have been compared to similar measurements on conventional surfactant-stabilized emulsions~\cite{Arditty05}. The interfacial elasticity dominates the behaviour in the case of colloid stabilization and this is consistent with the picture outlined in section III.A for flat interfaces. The most curious behaviour occurs on the scale of individual droplets which we detail next.

The crumpling of colloid-stabilized droplets (i.e. buckling of the interface) has been independently investigated by several groups: Masliyah and coworkers~\cite{Yeung99}~\cite{Dabros99} used a micropipette to probe the properties of a water droplet in a dispersion of bitumen in a heptane / toluene mixture. For concentrations of bitumen below 1\% volume the droplet crumpled as water was drawn out showing that the interface has become rigid under compression. Similar droplets were subjected to small volume-preserving distortions using two micropipettes and the recovery time was related to a model for a viscous interface. Because this system contains several constituents that are potentially surface active, analogous experiments were repeated using silica colloids on an interface between water and silicone oil~\cite{Asekomhe05}. Reversible crumpling was again observed, demonstrating that the solidification is associated with particles trapped by the interfacial tension. Quantitative studies in a similar vein have been carried out by Xu and coworkers~\cite{Xu05} who recorded the internal pressure of water droplets covered with polystyrene particles in a continuous decane phase, Fig~\ref{Crumpling}. They observed a discontinuity in the internal pressure at the onset of buckling for droplets of sizes above and below the capillary length. Using evaporation rather than suction to reduce the internal volume Abkarian and coworkers showed that faceting of armoured bubbles ultimately leads to stability~\cite{Abkarian07}. Attempts have also been made to raise the area fraction of interfacial colloids by causing the trapped nanocomposite microgel particles to swell. Changes of pH gave rise to swelling of the particles; however, this has led to a break-up of the emulsion~\cite{Fujii06}.

\begin{figure}[b]
\centerline{\includegraphics[scale=0.65]{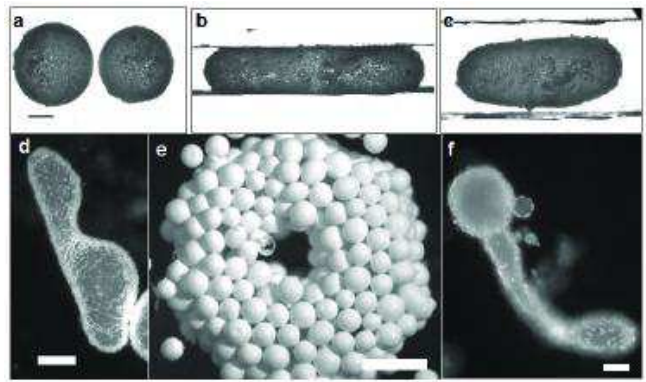}}
\caption{\label{donut_lite} Non-spherical bubbles and droplets (a) initial bubbles stabilized by 2.6~$\mu$m diameter polystyrene colloids. (b) Bubbles forced to fuse using cover slides and (c) a non-spherical shape is retained after the force is removed. (d \& e) Saddle and toroid shaped bubbles and (f) non-spherical oil droplets. Scale bars are: a-c, 100~$\mu$m; d, 200~$\mu$m; e, 500~$\mu$m; and f, 16~$\mu$m. Reprinted by permission from Macmillan Publishers Ltd: Nature~\cite{Subramaniam05}, copyright 2005.}
\end{figure}

Subramaniam and coworkers~\cite{Subramaniam05} have harnessed the concept of a solid interface to create non-spherical bubbles and droplets stabilized by colloidal particles (Fig.~\ref{donut_lite}). Air bubbles were trapped in packed beds of colloids and the resulting armoured bubbles were reshaped using a spatula; by this technique they were able to create interfaces with saddle curvature - including a toroid. These bubbles and droplets have interfaces that exhibit variations in mean curvature, which is only possible if either the fluid-fluid interface has close to zero interfacial tension or, alternatively, is solid. The latter applies in this case as is demonstrated by the static nature of the distortions. While the behaviour of this interface is solid-like on length scales larger than the colloids, it remains fluid in the interstices between colloids. They probed the mechanical characteristics more directly, first, by subjecting a bubble to shear flow and, second, by deforming bubbles using cover slides~\cite{Subramaniam06}. They concluded that the interfacial colloids have some similarities to a granular medium: they need to dilate in order to flow. Attempts to distort a complete droplet are typically met with an elastic response since a large stress is required to rearrange particles across the whole surface. By contrast a local stress leads to a plastic response since a small region can be strained to allow particles to slide past each other.

Another route to creating non-spherical surfaces is to fuse droplets in an emulsion of partially-miscible liquids stabilized by colloids~\cite{Clegg05}~\cite{Clegg07}. The emulsion can be created within the demixed region of the phase diagram via either nucleated phase separation or direct mixing. On warming, the droplets are quite stable well into the mixed region of the phase diagram, suggesting either that the colloid-laden interface presents a considerable impediment to diffusion, or demonstrating the role played by attractions between the colloids. All the same, the droplets become more likely to coalesce and the recently fused droplets are observed to maintain a peanut shape for many seconds, showing that variations in the mean curvature of the interface can still be supported on a large scale. Extended convoluted arrangements of domains can be created by warming droplets into the mixed phase and then quickly quenching the survivors back into the demixed regime~\cite{Clegg07}. The domains which form are irregular with a rough surface but with some indication that they are built from pre-existing droplets. The temperature dependent interfacial tension is being used to control the solidity of the interface and provides a useful mechanism for creating colloid-stabilized emulsions with novel geometries.

A cylindrical interface can be created via liquid jets and permanently stabilized using colloids~\cite{Edmond06}. Here a stream of one fluid (water containing polymer) was pushed through another flowing fluid (hexadecane) where the extensional viscosity was tuned to create a stable jet. When dispersed colloids were added to the hexadecane they adsorbed to the water-hexadecane interface. Particle-coated cylinders were created that remained stable after the flow had ceased, demonstrating that the interface is likely to be solid, at least on the length scale of the cylinder. This approach to creating stable cylinder-like domains will be returned to in section VI.B.

\section{Bicontinuous colloid-stabilized emulsions}

Demonstrations of the fabrication and properties of colloid-stabilized non-spherical droplets point the way towards the creation of more elaborate arrested states. In this section we discuss fluid-bicontinuous structures which also require variations in the mean curvature of the interface but which additionally exhibit novel macroscopic properties.

\subsection{Liquid-liquid phase separation} 

It is possible to create a bicontinuous arrangement of fluid domains in a variety of ways; however, the bijels comprised of low molecular-weight liquids reviewed here were all realised via liquid-liquid demixing~\cite{Rowlinson1982}. The demixing route will direct structure formation, and two distinct pathways to demixing are usually possible. For a quench that is deep and fast or for a composition that is close to a critical point the liquids demix via spinodal decomposition~\cite{Debenedetti96} where the mixed state of the pair of liquids becomes energetically unstable. At first the two phases diffuse apart~\cite{Bray94}~\cite{Kendon01}, and the interfaces are the same size as the domains themselves. With time the composition difference between the domains becomes more pronounced, the interfaces narrow, and interfacial tension (resisted by the fluid viscosity) begins to drive the phase separation. As the domains continue to coarsen the liquid motion is resisted by inertia. The alternative demixing route, nucleation, occurs for a quench that is shallow or slow or for a composition where the volume fractions of the two phases are very unequal~\cite{Debenedetti96}. Droplets of the minority phase tend to demix, and if they exceed a threshold size they continue to coarsen; otherwise they dissolve again. Both of these demixing routes will create liquid-liquid interfaces, although the initial spinodal interfaces will be broad and diffuse, while the initial nucleation interfaces will be narrow and distinct. Colloids which are dispersed in the mixed phase might become trapped on the freshly created interface as the liquids demix.

\begin{figure}
\centerline{\includegraphics[scale=0.65]{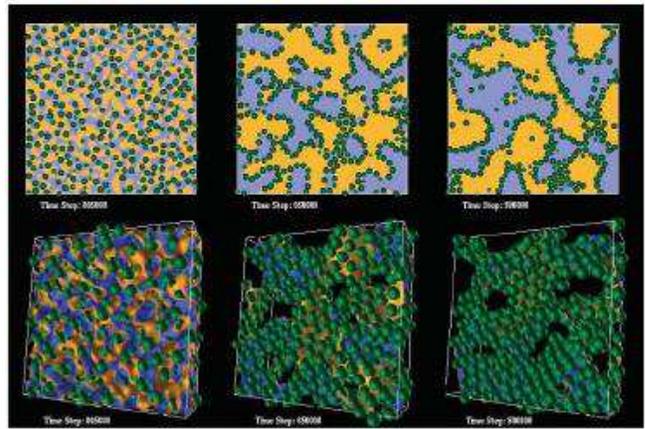}}
\caption{\label{simulations} Showing neutrally wetting colloids (green) at volume fraction 0.2 dispersed in a binary solvent that is quenched. The fluid domains (yellow and blue) evolve with time while the particles sequester to the interface between them. A more-or-less arrested bicontinuous structure is formed. From K.~Stratford \textit{et al} 2005 \textit{Science} \textbf{309} 2198. Reprinted with permission from AAAS.}
\end{figure}
Prior to interest in the formation of arrested states, significant research was carried out on the behaviour of dispersed particles in phase-separating liquids. Initial activity was due to Lash Miller and McPherson~\cite{Lash08} who investigated the partition of colloids between phase-separating liquids. Their studies of silver particles in water / phenol mixtures were abandoned because all silver precipitated, due to aggregation during phase separation. A body of data has subsequently built up associated with the technique of separating biological molecules using immiscible liquids~\cite{Albertsson60}. Later, Gallagher and Maher systematically explored the behaviour of polystyrene colloids of differing size and surface-charge densities in water-2,6-lutidine mixtures close to the critical point~\cite{Gallagher92}. The authors suggest that early in phase separation particles do not become trapped at interfaces because the interfacial tension is too low. Instead they are mainly found in a preferred phase. At some higher temperature a sizable interfacial population begins to appear which slows down phase separation. For a sufficiently large interfacial population they observed the formation of metastable foams with crystalline interfaces that were sometimes stable for days: this may have been the first bijel. Tanaka has explored how both mobile and immobile particles, with a preference for one of the fluids, influence the patterns formed by the domains~\cite{Tanaka94}~\cite{Tanaka01}. If the domains have a droplet morphology or the mobile particles are concentrated enough to crystallize then the time dependence of the domain growth is also modified. Computer simulations are currently being used to explore effective colloid-colloid interactions that may be induced at an early stage during phase separation~\cite{Araki06}. A range of experiments have probed the behaviour of dispersed particles prior to the onset of phase separation of the host solvent. These include very recent studies which show the influence of the critical Casimir force~\cite{Hertlein08} for a particle close to a flat surface. For non-critical compositions the studies of Beysens and others show pretransitional aggregation of colloids occurring when approaching the binodal from the single-fluid phase~\cite{Beysens99}~\cite{Koehler97}. 

\subsection{Bijels: computer simulations}

The initial explorations of the interplay between wetting and phase separation using computer simulations are due to Ginzburg and coworkers~\cite{Ginzburg99}~\cite{Ginzburg99b}. Here a multi-scale model was developed to show how particles, which are preferentially wetted by one of the fluids, influence the evolution of the phase transition. Ultimately a balance is achieved between the creation of new interfaces due to the motion of the particles and the coarsening of the domains. These and other early studies have been reviewed by Balazs~\cite{Balazs00}. The idea of modelling neutrally wetting particles in phase-separating fluids was first touched on by Tang and Ma~\cite{Tang02} although they did not encounter the arrest of domain growth. Laradji and coworkers have carried out detailed simulations of particles in phase-separating fluids in two and three-dimensions~\cite{Laradji03}~\cite{Laradji04}~\cite{Laradji04b}; for mobile particles with a preference for one of the fluids they find that it is predominantly the prefactor of the domain growth law that is modified which agrees with some of the experimental results of Tananka~\cite{Tanaka94}. 

The creation of fluid-bicontinuous emulsions stabilized by colloids via liquid-liquid demixing was also first investigated systematically using computer simulations~\cite{Stratford05}. The system consisted of hard-sphere colloids in a phase-separating liquid where the demixed phases were taken to have identical density, viscosity and volume fraction. Consequently the three-phase contact angle was $\theta_w = 90^{\circ}$. The lattice-Boltzmann technique was used to simulate the behaviour of the system as the liquids demixed via spinodal decomposition. Previous studies had explored the response of the unadulterated liquids to an instantaneous change in temperature~\cite{Kendon01}. The simulations show that in the presence of colloidal particles demixing continues to be observed; however, as the liquid-liquid interface forms it begins to affect the behaviour of the particles. The particles become trapped on the interface (see Fig.~\ref{simulations}) and as the domains coarsen they are forced together and eventually form a jammed monolayer~\cite{Stratford05}. It has been demonstrated for flat interfaces and droplets (sections III.A and III.B) that an interface densely coated with particles has semi-solid character, at least on macroscopic length scales. If this interface percolates through a sample, creating fluid-bicontinuous domains, then the sample should be a gel. To test whether a permanent gel was formed the separation between interfaces was studied for the duration of the simulation, and it was found that the speed of domain coarsening was drastically slowed down. Whether the system had fully arrested remains ambiguous as a result of the necessarily short duration of the simulation. Further higher-resolution simulations of cylindrical and wavy interfaces were carried out to demonstrate that the subsections of a bicontinuous pattern were indeed arrested. Because the simulations involved an instantaneous change in temperature~\cite{Carmesin86}, perfect hard-sphere colloids and perfectly symmetrical fluids, there is some distance between them and experiments. Nonetheless these simulations motivated the low molecular-weight liquid experiments presented here (sections IV.C and IV.D). Other computer simulations~\cite{Gonzalez-Segredo04} have been used to demonstrate that amphiphiles with no long-range interactions can arrest phase separation via the formation of an ordered state. It would be interesting to try to achieve this using colloids although it may require a very uniform wetting angle or amphiphilic (Janus) particles. Very recently Hore and Lardji have simulated high volume fractions of neutrally wetting particles in phase-separating fluids using a dissipative particle dynamics approach~\cite{Hore07}. They find that phase separation saturates at a finite domain size with an interfacial coating of particles and they demonstrate that this state is thermodynamically metastable. They explore this behaviour as a function of particle size and interfacial tension and show that, except in the case of extremely small particles, this state is likely to be very long lived. 

\subsection{Two-dimensional bijels: experiments} 

\begin{figure}
\centerline{\includegraphics[scale=0.65]{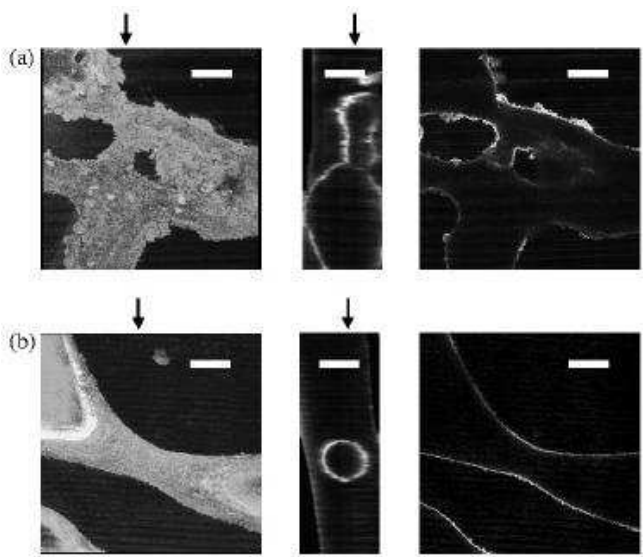}}
\caption{\label{thin} (a) Left: Rendered confocal images of structures viewed from above. Centre: vertical slices corresponding to the position of the arrow on the left. Right: horizontal slice through the confocal stack on the left at the position of the central arrow (b) An isolated fluid neck, ending at contact lines with the sample cell (Left, centre and right as in (a)). There are 2\% volume of 440~nm diameter silica colloids and 38.7:61.3 methanol:hexane by volume. The fabrication procedure is described in the text. Scale bars are 100~$\mu$m. Reprinted with permission from~\cite{Clegg07}. Copyright 2007 American Chemical Society.}
\end{figure}
Experiments focused on the alcohol-alkane combinations methanol-hexane and ethanol-dodecane~\cite{Clegg05}~\cite{Clegg07}, and via careful control of the particle surface chemistry, sample composition and cooling rate it was possible to arrest thin bicontinuous domain arrangements. Silica colloids were used and to achieve partial wettability a silanization procedure was employed~\cite{Horozov03}. The particles (typically 2\% volume fraction) were dispersed at a temperature where the two fluids were miscible. The composition of the liquids was chosen to be half-way between critical composition and equal volumes of the two liquids; the former favours spinodal decomposition while the latter facilitates a bicontinuous arrangement of fluid domains geometrically. On cooling, the three-component sample passes through a region of the phase diagram where nucleation is the preferred route to demixing. Since we aim to make use of spinodal decomposition it is essential that nucleation should be suppressed. We achieved this by cooling the thin samples quickly in baths of cryogens. The domain arrangements were viewed using confocal microscopy (Fig.~\ref{thin}): the fluid domains appear black while the colloid-laden interfaces appear white. An uncoated fluid cylinder, a basic subunit of a bicontinuous pattern, immersed in another fluid will tend to break up via the Rayleigh-Plateau instability. This is because the interfacial area can be lowered by the cylinder evolving into droplets. As can be seen (Fig.~\ref{thin}) the interfacial layer is able to suppress this effect consistent with it being semi-solid. In other frames it can be seen that there are variations in the mean curvature on a large scale. This demonstrates that the interfaces have, to some extent, solidified. Constant mean curvature of the liquid-liquid interface is maintained in the interstices.

These results appear to confirm the predictions of the computer simulations~\cite{Stratford05}; however, in a number of important respects they are substantially different. Firstly, the layer of colloids coating the interface is much more than a monolayer; this is evident in Fig.~\ref{thin} where the particle layer can be seen to be quite fluffy. By comparing frames it is evident that this feature is quite sensitive to the details of the quench route~\cite{Clegg07}. It is likely that the thickness of the interfaces is a consequence of the residual interactions between the particles. Another unavoidable difference between the experiments and the computer simulations is that the former were carried out at a finite (albeit fast) quench rate with the heat being removed via the surface. It was found that there was an optimum cooling rate for the samples and this was not the fastest possible. Pre-cooling in liquid nitrogen prior to quenching in either ice or dry ice led to structures with the clearest resemblance to a spinodal pattern (cooling directly in a dry ice bath would lead to a faster quench). It is possible that severe cooling of the surface may lead to a wetting effect previously studied by Tananka~\cite{Tanaka96} where flow along the domains that form during spinodal decomposition greatly speeds the creation of a thick wetting layer. This also leads to the preferential growth of cylindrical domains perpendicular to the surface and hence disrupts the formation of the characteristic spinodal pattern. The second way in which the cooling rate affected our results was via the thickness of samples that could be created. We were able to make stable structures with thicknesses of 200~$\mu$m as seen in Fig.~\ref{thin}, but attempts to make samples twice this thickness were not fully successful: the sample and the structure reverts to droplets towards the centre. This suggests that the heat was not being removed from the centre of the sample quickly enough; hence this sample composition is not ideal for studying fully three-dimensional structures. To pursue these goals we moved on to a different system.

\subsection{Three-dimensional bijels: experiments} 

Adjustments to our experimental protocol led to the creation of fully three-dimensional bicontinuous gels~\cite{Herzig07}. We used water-lutidine which has a phase diagram with greater symmetry than that of the alcohol-oil combinations described above and this facilitates quenching through the critical point. St\"{o}ber silica tagged with fluorophore FITC exhibits partial wettability towards these two solvents. The cuvette (1~$\times$~10~$\times$~50~mm$^3$) was introduced into a block of aluminium that was preheated to 40~$^{\circ}$C giving a warming rate of 17~$^{\circ}$C/min. The block was employed because it was found that temperature gradients impeded the formation of a bicontinuous gel structure. It was sited on an inverted confocal microscope, and demixing in the presence of the colloids could be observed in real time, showing that during demixing via spinodal decomposition the interface between the two domains becomes brighter as the domains coarsen~\cite{Herzig07}. It can be seen that spinodal decomposition is being arrested by the presence of the interfacial layer of particles.
\begin{figure}
\centerline{\includegraphics[scale=0.65]{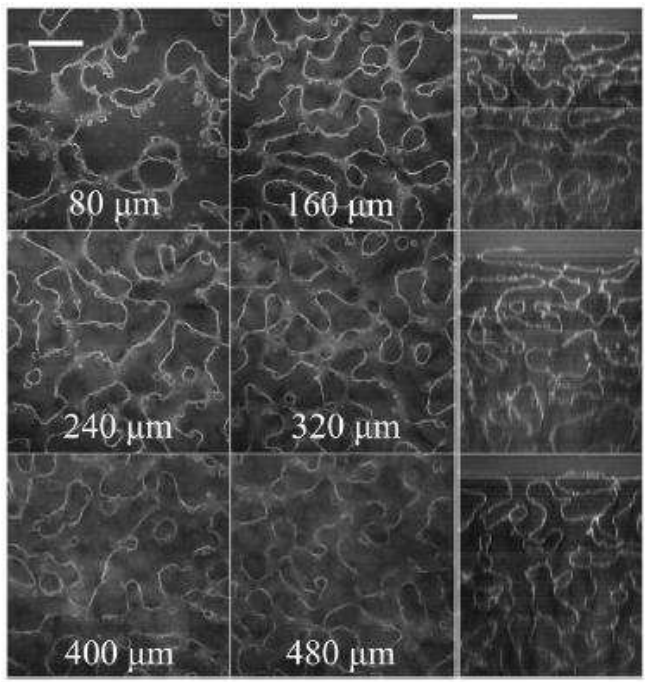}}
\caption{\label{thick} Left: fluorescence confocal microscopy images at different depths into the sample of 2,6-lutidine-water at critical composition with 2\% volume fraction of silica colloids quenched from room temperature to 40$^{\circ}$C at 17$^{\circ}$C/min. Depth in the sample is shown on the labels. Right: reconstruction along the vertical axis (thinnest dimension) for bottom, centre and top images on the left reaching 500~$\mu$m into the cuvette. The scale bars are 100~$\mu$m. Reprinted by permission from Macmillan Publishers Ltd: Nature Materials~\cite{Herzig07}, copyright 2007.}
\end{figure}

Figure~\ref{thick} and associated movies~\cite{Herzig07} show a  stack of images (411~$\times$~411~$\times$~500~$\mu$m$^3$) that is sufficiently thick that numerous fluid domains are encountered from one side to another, indicating that this is a fully three-dimensional bijel. The structure is reminiscent of a sponge phase. However the interfaces of a sponge phase have vanishing interfacial tension and fluctuate strongly to the extent that they are constantly breaking and reforming~\cite{Adelman87}. Here the structure is entirely static and when stored at 40$^{\circ}$C for many months no ageing behaviour is observed, strongly suggesting the interfaces are solid, at least on the macroscale. The solidity has been tested in a provisional manner by observing the response of the system to a moving heavy object. The arrangement of interfaces is able to support the weight of a 200~$\mu$m diameter copper wire that weighs 1.9~mg. When dropped into the gel the wire travels a short distance as it slows down to a stop. This is consistent with the bijel being a viscoelastic medium characterized by a particular relaxation time and yield stress. The relationship between the fluid-bicontinuous gels that we create and the underlying phase diagram can be usefully compared with catastrophic inversion~\cite{Binks00c} (see section II.C). We demonstrated this by preparing a sequence of colloid-stabilized emulsions using the same volume fraction of particles and warming rates but with different compositions of fluids. When the critical composition is employed a bijel forms, while for compositions either side of critical the minority phase became the dispersed phase for a droplet emulsion.

The nature and formation of the arrested state can also be probed by changing the colloid volume fraction (here we used particle volume fractions from 0.5\%$_v$ to 4.0\%$_v$). All samples have critical composition and were warmed at 17$^{\circ}$C/min to 40$^{\circ}$C, and as the volume fraction of particles increases the separation between the colloid-coated interfaces decreases. This is purely a geometrical effect, resulting from the amount of interface required to accommodate the particles. Hence the sample with the most particles has arrested at an earlier stage during spinodal decomposition, confirming our picture of the sweeping up and subsequent jamming of particles. For the smallest volume fractions it appears that the spinodal structure was not fully stable and some domains have broken up into droplets. It is possible that the particles are favouring a certain minimum curvature. The scaling of the domain size $\xi$ with particle volume fraction $\phi_v$ is described by $\xi \sim d / \phi_v$ where $d$ is the colloid diameter and the prefactor depends on the geometry of the interface and the thickness of the colloidal layer. High-resolution confocal microscopy provides confirmation that the interfaces are stabilized by roughly a monolayer of colloids~\cite{Herzig07}.

\section{Bicontinuous polymer blends stabilized by interfacial particles}

Independently, polymer blend researchers have established routes to creating bicontinuous structures stabilized by interfacial colloids. Bicontinuous domain arrangements have been achieved both with and without the use of polymer-polymer phase separation, while the particles used are sometimes complex.

\subsection{Polymer blends and particulate compatibilizers}

When polymers are blended the constituents would ideally be fully miscible; however, this is rarely the case and the aim is then to reduce the domain size down to a scale where the system is transparent and exhibits a single glass transition temperature~\cite{Rudin99}. Furthermore, it is very unusual to find pairs of chemically dissimilar polymers that are even partially miscible because the individual molecules are large, which reduces the entropy of mixing~\cite{Larson99}. This closes off one route to the formation of bicontinuous domain arrangements since access to phase-separation kinetics is only possible in a very small number of cases. Fortunately it has been found that formation of a bicontinuous structure via melt mixing or phase inversion is much more straightforward for polymers (section V.C) as compared to low molecular-weight liquids~\cite{Paul80}. Another response to immiscibility is to mix the pair of polymers in a solvent where equilibrium is more easily reached for dilute solutions, and the challenge is to find good solvents for both constituents.  

In polymer blending the interface is crucial because the stability of a composite can be undermined if there is poor adhesion between the constituents. The width of polymer-polymer interfaces (often referred to as the interphase~\cite{Ajji96}) tends to be a few nanometers, which is large compared to interfaces between low molecular-weight fluids~\cite{Jones97}. Helfand-Tagami lattice theory predicts that there will be an elevated concentration of chain ends of both polymers at the interface~\cite{Helfand71}~\cite{Ajji96}~\cite{Lipatov02} and that any low molecular weight third component will tend to collect there. Typically the blend morphology involves closed interfaces around a dispersed phase; however, there are indications that the formation of a bicontinuous structure will lead to a dramatic improvement in toughness without substantially altering other key engineering properties~\cite{Potschke03}. 

A common route to improving the compatibility of two polymers is to use additives, and there are many to choose from~\cite{Dutta96}. In this review, we are specifically interested in particulate additives and these may well play a similar role to the colloids in low molecular-weight bijels. Comparing emulsions of low molecular-weight liquids with blends of polymers is contentious because of the very different length and time scales involved. The range of depletion interactions between particles will be of the same order as the fluid subunits making this a serious issue for polymers. Demixing occurs in seconds for low molecular weight liquids while it can take place over days for a polymer blend: The interfacial tensions are similar for phase-separating low and high molecular-weight mixtures ($\sim$1~mNm$^{-1}$) however the viscosities can differ by five to six orders of magnitude (e.g. 10$^{-3}$~Pa~s for water-lutidine~\cite{Herzig07} and 10$^{3}$~Pa~s for a blend of poly(methyl methacrylate) and poly(styrene-\textit{ran}-acrylonitrile)~\cite{Chung07}). It is worth considering how this disparity in viscosities will manifest itself. The time scale associated with colloid motion is $\tau_R \sim 6\pi \eta R^3 / k_B T$: the time for a particle to diffuse a distance equal to its own radius and the time taken for the liquid-liquid interface to move a similar distance is $\tau_I \sim R \eta / \gamma$. Both of these scale linearly with the viscosity implying that all motion will slow down in the same way. Since it is not clear at what point during separation the colloids become trapped on the interface this argument could well be misleading. If the particles become trapped at the early (diffusive) stage this scaling analysis does not hold~\cite{Bray94}. If the particles become trapped at a late stage there are a variety of considerations associated with hydrodynamic flow and film thinning which will all change with the viscosity of the fluids~\cite{Nguyen06b}. The results presented in sections V.B and V.E have sufficient similarity to the low molecular-weight liquid results (section IV) to suggest that the most crude analysis may have some merit.

Vermant and coworkers~\cite{Vermant04} used silica nanoparticles to stabilize blends of polydimethylsiloxane and polyisobutylene. Stable droplets were found with the particles largely sequestered to the interface. Rheological studies suggest that break-up and coalescence are suppressed by the interfacial layer. This supports the idea that the same physics is important for polymer blends as for the low molecular-weight emulsions. By contrast, for diblock copolymer compatibilizers it is normally assumed that the interfacial tension of the two constituents is being reduced i.e. the diblock copolymer plays the same role as a conventional surfactant (in at least one case it has been suggested that coalescence of the dispersed phase is suppressed by steric hindrance~\cite{Jones97}). The idea that some compatibilizers reduce the interfacial tension is strongly supported by the studies of Bates and coworkers~\cite{Bates97}~\cite{Fredrickson97}. Here the equilibrium phase diagram of three component: polymer-polymer-diblock copolymer blends were studied. A bicontinuous microemulsion was found in a narrow window of composition between the lamellar and two-phase regions of the phase diagram: this is a state where the interfacial tension between the two polymers is very low. The studies were repeated with unusually low molecular-weight polymers to confirm that the mesophases were in equilibrium~\cite{Hillmyer99}.

In a related direction that was first explored theoretically~\cite{Thompson01}~\cite{Lee03}, the surface chemistry of gold nanoparticles has been tuned so that they preferentially occupy the interfaces of a diblock copolymer sample~\cite{Chiu05}~\cite{Kim06}. The equilibrium state of the diblock copolymer, prior to the addition of particles, is a lameller phase and this has large quantities of interface, albeit flat. The addition of neutrally wetting particles at sufficiently high concentrations destroys the lamellar phase in favour of a bicontinuous domain morphology~\cite{Kim07}. The authors suggest, supported by theory~\cite{Pryamitsyn06}, that this indicates that particles at polymer-polymer interfaces reduce the interfacial tension and subsequently the bending modulus between the diblocks. It is also possible that the nanoparticles lead to the break-up of the lamellar phase by introducing and pinning defects in the ordering of the lamellae. A layered arrangement with short-range order might look similar to structures observed in ref.~\cite{Kim07}.

\subsection{Two-dimensional bijels via spin-casting}

Spin casting a partially miscible blend, initially dissolved in a common solvent, has been thoroughly explored as a way to create a spinodal pattern. Composto and coworkers began by probing the phase separation of thin layers of polymer blends~\cite{Wang00}; they went on to look at blends containing particles that partition into one of the phases~\cite{Chung04} and most recently blends containing neutrally wetting particles~\cite{Chung05}. Due to the intrinsic slowness of polymer blend phase separation, they are able to investigate changes to kinetics due to the influence of particles. Their studies are carried out using a range of scattering and scanning microscopy techniques with blend film thicknesses ranging between 100~nm and 1500~nm. The polymer blend, deuterated PMMA (dPMMA) and poly(styrene-\textit{ran}-acrylonitrile) (SAN), has a symmetric phase diagram and a lower critical point for composition $c \sim 0.5$. It is prepared in a solvent (methyl isoamyl ketone), spun cast and then dried for 24~hrs. On warming into the demixed regime four stages of separation are seen. This is a modification to standard spinodal decomposition, described in section IV.A, due to the dominance of the free and substrate surfaces in these samples. First, the pre-stage lasts about two minutes and is presumably dominated by diffusive phase separation. Second, the early stage lasts two hours; here wetting layers of dPMMA are established at the free and substrate surfaces. The mid section of the film separates via a bicontinuous morphology. Hydrodynamic flow from this region leads to the rapid growth of the wetting layers, leading to a trilayer structure. Third, the intermediate stage lasts 48~hours: the dPMMA flows back from the wetting layer and forms columns that span the SAN-rich mid-layer connecting the two wetting layers. Finally, in the late stage, the SAN-rich layer ruptures and becomes droplets within a dPMMA continuous region. The stages of phase separation are manifest in the domain morphology and also in the roughness of the free surface of the film.
\begin{figure}
\centerline{\includegraphics[scale=0.65]{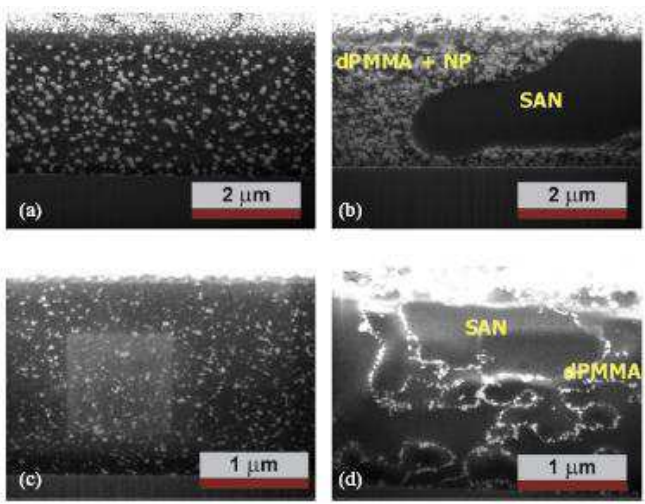}}
\caption{\label{poly-thin} Cross-sectional SEM images of dPMMA:SAN films with 10 wt \% MST (a \& b) and 10 wt \% P2K (c \& d). Initially, MST (a) and P2K (c) are homogeneously dispersed. Upon annealing at 195$^{\circ}$C for 24 h, MST nanoparticles partition into the dPMMA phase (b) whereas P2K nanoparticles segregate at the dPMMA/SAN interface (d). Reprinted with permission from~\cite{Chung05}. Copyright 2005 American Chemical Society.}
\end{figure}

The relationship between the particle wettability and the behaviour of the fluid-fluid-particle composite is quite familiar. The first three-component system to be studied involved nanoparticles (MIBK-ST, Nissan Chemical referred to as MST) that preferentially partition into the PMMA polymer~\cite{Chung04}. In the single-phase region they are homogeneously mixed. The role of the nanoparticles is to enhance the viscosity of the PMMA and this slows phase separation (quite substantially for 10\% of particles). The second study~\cite{Chung05} was of 5\% and 10\% weight of nanoparticles (DMAC-ST, Nissan Chemical with grafted PMMA with a chlorine end group referred to as P2K) which have similar wetting character with the two polymers and hence prefer the interface (equation~(\ref{Energy})). The arrangement of the samples after 24~hours is studied when blend samples would be in the intermediate stage of phase separation. The interfacial particles permanently arrest the polymer demixing prior to the mid-section SAN film rupturing (see Fig.~\ref{poly-thin}); consequently, the bicontinuous morphology is retained. It is impressive that the same stabilization mechanism appears to perform successfully in such a physically different composite to the low molecular-weight bijels (see sections IV.C and IV.D). 

The influence of neutrally wetting particles is also visible externally. When particles are absent the film surface shows a capillary instability and roughness which have a related time dependence. In studies of the interfacial roughness of the film the particles that partition into one of the polymers have little influence~\cite{Chung07}. Interfacial nanoparticles suppress both the capillary instability and interfacial roughening~\cite{Chung05}. This is presumably associated with the stabilization of the SAN mid-section of the trilayer. Minelli and coworkers have also used spin casting to organize nanoparticles in polymer blends~\cite{Minelli04}~\cite{Minelli06}. In this case particles with a preference for one of the polymers are considered (although some enrichment at the interface is observed). They have investigated the influence of patterning the surface that the blend is spun onto and also the domain size, thickness and roughness of the film.

\subsection{Bicontinuous polymer domains via phase inversion and melt mixing}

The conditions for bicontinuity have been considered for an arbitrary domain geometry by Onuki~\cite{Onuki02} and in terms of the stability of fluid cylinders by Willemse~\cite{Willemse99}. The estimated composition for a bicontinuous configuration of arbitrary geometry is
\begin{equation}
\frac{\phi_1}{\eta_1} \approx \frac{\phi_2}{\eta_2}.
\label{Onuki}
\end{equation}

\noindent A low-viscosity, $\eta$, component will tend to become continuous during mixing to reduce dissipation. To counteract this the volume fraction, $\phi$, of a higher-viscosity component must be increased~\cite{Onuki02}. Equation~(\ref{Onuki}) tends to describe only the behaviour of experimental systems when the viscosity ratio is close to unity. Bicontinuity is often characterized by the formation of interconnected, roughly cylindrical domains~\cite{Potschke03} and this can be used to explore the likely formation conditions. The stability of droplets and fluid cylinders under shear is described by the capillary number
\begin{equation}
Ca = \frac{\hbox{shear stress}}{\hbox{Laplace pressure}} = \frac{\eta_m \dot{\gamma} R_0}{\sigma}
\end{equation}

\noindent where $\sigma$ is the interfacial tension, $\eta_m$ is the viscosity of the continuous matrix, $\dot{\gamma}$ is the shear rate and $R_0$ is the radius of a domain if it were spherical. For values of Ca above a critical threshold droplets break-up under the influence of shear~\cite{Larson99}. The droplets deform before they rupture and the creation of long cylinders is possible using a sudden increase in flow rate. Cylinders might also be formed at high shear rates if the system has a highly viscous continuous matrix and a low interfacial tension and if the initial droplets are large~\cite{Potschke03}. Additionally, break-up can be inhibited by having the cylinders shorter than the dominant wavelength of the capillary wave instability, as explored by Willemse~\cite{Willemse99}. Stability via this condition occurs over a broader composition range for blends when the zero shear-rate viscosity ratio is substantially different from one. The addition of filler particles to a blend can help in this regard and hence lead to a substantial slow-down in the rearrangement of bicontinuous domains~\cite{Potschke03}. For some blend / filler combinations the enhanced stability is due to a change in the viscosity of one of the constituents; however, in other cases cylinders can also be formed that exhibit a yield stress.

The two ways to create a bicontinuous domain arrangement for immiscible polymers are phase inversion (which relies on variations in polymer viscosity) or chaotic mixing of the melt~\cite{Paul80}~\cite{Potschke03}. Emulsion inversion has been discussed in the context of low molecular-weight liquids (section II.C) where the inversion point is unstable when approached by direct mixing. The polymer phase inversion occurs because the blend fulfils the condition described by equation~(\ref{Onuki}) during mixing as a consequence of the dependence of viscosity on shear rate, temperature or chemical reactions. During inversion a bicontinuous structure is created which is stable if the cylinder stability conditions are met. The range of compositions for which this is true has been determined by Willemse and coworkers~\cite{Willemse98}. Bicontinuous structures can also be stabilized by prompt cooling following mixing. Chaotic mixing of a melt can lead to bicontinuity due to the repeated stretching and folding of polymer sheets~\cite{Potschke03}~\cite{Aref02}. As this procedure continues the sheets of the low viscosity minority constituent become thin enough to rupture leading to an interconnected arrangement of domains. This is fundamentally different to phase inversion since this process does not rely on a change in viscosity to take the system through the condition for bicontinuity: equation~(\ref{Onuki}). Melt mixing creates a domain arrangement that is bicontinuous in a system that already fulfils this criterion.

\subsection{Clay at polymer-polymer interfaces}

\begin{figure}
\centerline{\includegraphics[scale=0.8]{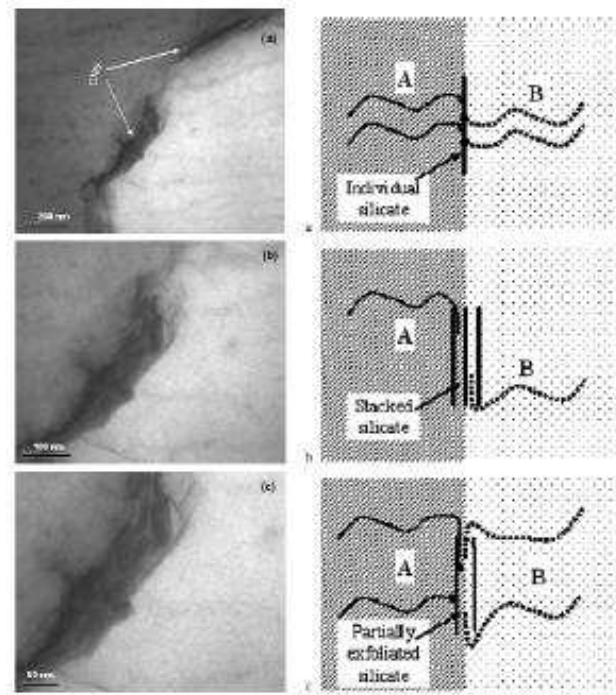}}
\caption{\label{clay} Left: Bright-field TEM images of the interface of 17.5 polystyrene / 77.5 polypropylene blend compatibilized by 5C20A Cloisite organically modified clay. Reprinted with permission from~\cite{Ray04}. Copyright 2004 Elsevier Limited. Right: schematic of possible compatibilization mechanisms for layered silicates. Polymers do not have to interact directly with the clay, they may just become trapped in the clay gallery. Copyright 2007 Society of Plastics Engineers. Reprinted from~\cite{Fang07} with permission of John Wiley and Sons.}
\end{figure}

The final experimental results we will consider concern melt-mixed polymers stabilized by interfacial clay platelets and these particles require an introduction. The dispersal of clay in homopolymers took off in the late 1980s with Toyota's exploration of the exfoliation of clay in nylon-6~\cite{Gao04} which led to a significant improvement in a wide range of engineering properties. These composite materials now feature in car parts and drinks packaging, but the mechanisms by which the clay enhances the polymer properties are not yet fully understood. Typically smectite clays are used which are comprised of galleries of silicate layers each about 1~nm thick. As a pretreatment it is common to make the hydrophilic clay hydrophobic, sometimes using surfactants, and the resulting particles are called organoclays. The clay platelets are several hundred nanometers along the long edges, Fig.~\ref{clay}a.

Recently, studies have begun to probe the influence of organoclays on the properties of polymer blends, and we focus on where the particles are located in different polymer pairs. Wang and coworkers studied montmorillonite (MMT) clay in blends of polypropylene (PP) and polystyrene (PS)~\cite{Wang03}. For organo-MMT in 70:30 blends of PP/PS the dispersed domain size decreased systematically with the volume fraction of clay and with the duration of mixing. X-ray scattering reveals that polymer (probably of both kinds) is likely to be intercalated into the clay galleries. They suggest that the clay decorated with polymer is likely to be at the interfaces of domains. Ray and coworkers go further~\cite{Ray04} and provide evidence of the presence of clay at polymer-polymer interfaces (Fig.~\ref{clay}a). Using an organically modified montmorillonite clay (Cloisite) they see dispersed-phase droplets decreasing in size with increasing clay concentration. High-resolution TEM images reveal that intercalated clay layers are trapped at the interfaces between polymer regions. They suggest that this arrangement is lowering the interfacial tension between the polymers. The presence of clay at the interfaces was also shown for blends of poly(butylene terephthalate) (PBT) and polyethylene (PE) using Nanofil (modified montmorillonite)~\cite{Hong07}.

Commercial clays contain several constituents which can complicate the identification of a stabilization mechanism. Ray and Bousmina looked at a range of clays (all MMT) in blends of polycarbonate (PC) and PMMA and by using electron microscopy studies, showed that the improved dispersion was not solely due to the cation used to treat the clay~\cite{Ray05}. Improved compatibilization was achieved by clays with larger spacing between layers, and they suggest this may be associated with the clay gaining a coating of mixed polymer. The presence of an adsorbed polymer coating was shown another way by Su and coworkers who exposed the interface between the two polymers PP and polyamide 6 (PA6) by cooling the samples and dissolving the PA6 with formic acid~\cite{Su07}. Using TEM of blends, modified montmorillonite clay was observed to sequester to the interface. The PA6 could be completely removed if no clay was present while a remnant quantity of PA6 was associated with interfacial clay. Fang and coworkers studied the morphology of PA6/high density polyethylene (HDPE) and PA6/HDPE-\textit{graft}-acrylic acid (PEAA) blends and the location of added clay (Oclay: C18-bentonite)~\cite{Fang07}. They found that a large proportion of the clay was still stacked in plates and was located at the polymer-polymer interface. They suggest a picture for the clay in which both species are intercalated between the platelets (Fig.~\ref{clay}b); they imagine this will occur so as to configure the composite particles to look like large block copolymer slabs. However the results of the studies of dissolving PA6 with formic acid~\cite{Su07} appear to suggest that both species could be randomly distributed around the clay surface.

\subsection{Three-dimensional bijels from polymer blends}

\begin{figure}
\centerline{\includegraphics[scale=0.65]{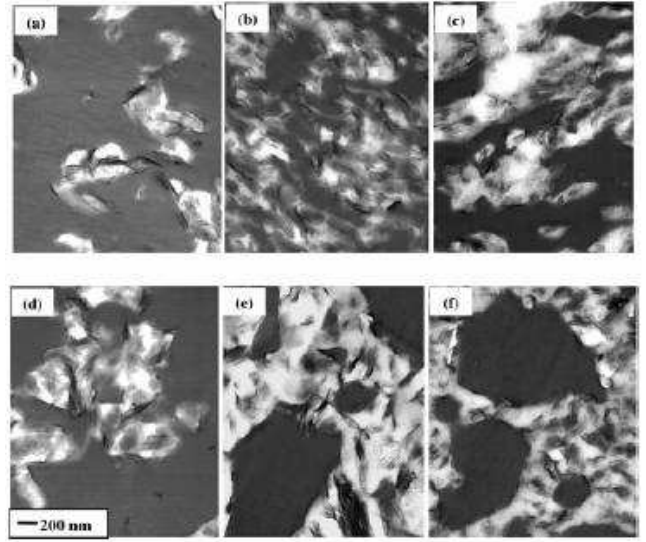}}
\caption{\label{Rafailovich} PS/PMMA/Cloisite 6A (10\%) blends quenched in liquid nitrogen shown using high-magnification TEM. The compositions are: (a) PS/PMMA (63/27), (b) PS/PMMA (45/45), (c) PS/PMMA (27/63). Images after annealing at 190$^{\circ}$C for 14~h for: (d) PS/PMMA (63/27), (e) PS/PMMA (45/45), (f) PS/PMMA (27/63). Reprinted with permission from~\cite{Si06}. Copyright 2006 American Chemical Society.}
\end{figure}
The clay compatibilizer studies lead towards the work of Si and coworkers who have explored the use of clays in the compatibilization of polymer blends of different domain geometries~\cite{Si06}. Unlike the systems used by Composto and coworkers~\cite{Chung05} the blends chosen, PS/PMMA and PC/SAN stabilized by MMT organoclays, do not require a solvent and the samples are fully three-dimensional rather than spun-cast films. Initial studies looked at the behaviour of the clay in each of the polymers individually. These showed that the clay platelets intercalated or exfoliated in the polymer; on extrusion the platelets have a common orientation with a long axis along the extrusion direction. Subsequently composites were prepared by mixing pairs of polymers at high temperature then adding the clay and further mixing. On cooling the 50~g samples were sliced and examined using TEM and scanning transmission X-ray microscopy (STXM). The polymer domains varied from droplets of PMMA, Fig.~\ref{Rafailovich}(a), through to droplets of PS, Fig.~\ref{Rafailovich}(c), in keeping with the change in the minority phase. In between, the system passed through a bicontinuous arrangement of domains. Here the clay platelets appear to be associated with PMMA domains although the results are not completely clear, Fig.~\ref{Rafailovich}(b). What is evident is that the platelets are no longer aligned along the extrusion direction. When the samples were further annealed the domain sizes increased and the platelets are seen to be creating faceted domain boundaries due to their location at the interfaces, Fig.~\ref{Rafailovich}(e). The polymer domain sizes investigated are in the range 100-1000~nm. An increase in the storage modulus of a factor of 2 is observed, while in some cases the composite exhibits only a single glass transition temperature. This appears to be a promising route for enhancing material properties.

The bicontinuous blends stabilized by clay platelets have a close resemblance to a bijel: there is a fluid-bicontinuous arrangement of domains stabilized by particles trapped at the interface. The authors describe the results in terms of polymer adsorbed to the clay surface being the key effect~\cite{Si06}. However, it is possible that this \textit{grafting} to the surface of the platelet occurs for both polymer varieties in the mixture resulting in the creation of partially wetting particles in situ. Motivated by recent experiments, theoretical attempts to model the compatibilization effect have begun~\cite{He06}. The model focuses on the changes to the free energy of the homogeneous blend due to the presence of particles. For the regime studied, the particles can have enthalpically unfavourable interactions with the polymers yet, in spite of this, the system does form a stable single-phase composite for a wide range of parameters.

\section{Future directions}

Based on current results we briefly outline potential bijel applications and then consider other fabrication routes especially in the middle ground between low and high molecular-weight systems. Bijels have been fabricated using low molecular-weight liquids building on previous studies of colloid-stabilized emulsion inversion and colloid-stabilized emulsions with non-spherical droplets. Very similar structures have been stabilized in polymer mixtures using either nanoparticles or clay platelets as the compatibilizers. Polymer blends with compatibilizers are already broadly applied in industry~\cite{Utracki98} where properties (including cost) that are intermediate between the two constituents are achieved by creating a stable blend. There are indications that a bicontinuous blend will lead to an enhancement of mechanical properties~\cite{Potschke03}. In addition to generic blend applications polymer-nanoparticle composites may well be valuable for photovoltaic films, fuel cells and batteries~\cite{Sivula06}~\cite{Ulbricht06}~\cite{Saunders07}. Low molecular-weight bijels have a substantial shared interface between two immiscible (or partially miscible) fluids~\cite{Dinsmore02} which could be valuable for controlled release~\cite{Lawrence07}~\cite{Simovic07}. There are many other applications for which it is useful to bring two immiscible fluids into contact: for example solvent extraction~\cite{Thayer05}~\cite{Gabelman99}~\cite{Kentish01}, catalysis~\cite{Vos82}~\cite{Krishna94}~\cite{Kiwi-Minsker05} and analysis~\cite{Jonsson01}~\cite{Smith03}. Here the fact that the fluid domains both percolate through the sample, allowing flow, is key.  

\subsection{Low molecular-weight liquid routes}

Currently bijels made from low molecular-weight liquids are limited to partially miscible pairs that phase separate into domains of roughly equal volume. One way to extend this range is to take advantage of the intrinsic properties of the structure. A bijel, like a bicontinuous microemulsion, is stable in the presence of an excess of either of the fluids since the domain of the minority fluid is protected by interfacial colloids. This means that it is possible to attach tubes and to flow liquids through the channels, and this could be used to retrospectively change fluids. The bijel should remain stable provided the new liquid has a high interfacial tension with the fluid that remains in the minority channel. Aspects of this approach have been explored via computer simulations~\cite{Stratford05} and ongoing experiments.

A recent discovery appears to have brought particle-stabilized emulsions and microemulsions closer together. This may point the way to a new route to fabricate bijels. Philipse and coworkers have created colloid-stabilized emulsions which appear to be energetically stable~\cite{Sacanna07a} and hence resemble microemulsions. These droplet emulsions are stabilized by nanoparticles (a broad range of different types of nanoparticles are effective); the oil is specifically 2-methacryloxypropyl trimethyoxysilane (TPM) and the continuous phase is water. The emulsions appear to be stabilized by the combined effects of the standard interfacial tension effect (equation~(\ref{Energy})), the line-tension contribution from these small particles~\cite{Aveyard03b}, double-layer repulsion between the absorbed colloids and the effect of the surface layer of TPM being hydrolysed~\cite{Sacanna07b}. The resulting emulsions are monodisperse and mixing two different droplet size emulsions results in monodisperse emulsion with an intermediate droplet size. The phase behaviour remains to be explored.

\subsection{Routes via blending properties}

It would be of great value to be able to fabricate a bijel of low molecular-weight liquids via direct mixing, and one way to think about this is to consider the effect of shear on phase-separating liquids. An imposed shear will effectively remix binary fluids at an early stage during spinodal decomposition and can be described in terms of a shift in the critical temperature of the system. This has been studied for low molecular-weight liquids by Beysens and coworkers~\cite{Beysens79}~\cite{Beysens83}. They found that the shift in critical temperature was well described by $\Delta T_c \simeq 2 \times 10^{-2} \dot{\gamma}^{0.53}$ where $\dot{\gamma}$ is the shear rate. A theory, due to Onuki and Kawasaki~\cite{Onuki02}, applies to the case of constant shear rate (which was not achieved in the experiments). The transition temperature shift is then $\Delta T_c \propto \dot{\gamma}^{1/3\nu}$ where $\nu = 0.630$ the correlation length critical exponent. The exponent agrees well with experiments, while the theoretical prefactor is four times too large. For low molecular-weight liquids the shift in $T_C$ amounts to several milli-Kelvin and so does not greatly expand the range of compositions under which spinodal decomposition can be achieved~\cite{Beysens83}. However, for other systems where the viscosity and the bare correlation length are substantially larger and/or the interfacial tension is substantially lower, this can be a significant effect. Pertinent systems include pairs of polymers in a common solvent~\cite{Larson92} and microemulsions that phase separate into micellar gas and liquid phases~\cite{Roux86}. In the polymer case transition temperature shifts of 10$^{\circ}$C are readily achievable. Hence for these kinds of system spinodal decomposition can be achieved by direct mixing at fixed temperature. 

Complex immiscible fluids have been created that include both molecular weight groups and may have properties which favour the creation of bicontinuous configurations. By dissolving polymers in water Edmond and coworkers~\cite{Edmond06} were able to create stable cylinders by flowing water in a flowing continuous phase of hexadecane. The cylinders are the basic elements of a bicontinuous configuration and the stability was due to the high extensional viscosity. Arditty and coworkers~\cite{Arditty05} also dissolved polymers in water; this time the water was emulsified in polydimethylsiloxane (PDMS). This facilitates tuning the system to a capillary number close to one. By tuning the properties of a system in this way it may be possible to create a low molecular-weight liquid bijel via chaotic mixing (section V.C). A similar approach is already used in polymer blending, where filler particles are used to modify the behaviour of phases, thereby expanding the regime of stability of a bicontinuous domain arrangement~\cite{Potschke03}.

In summary, we have reviewed the stabilization of interfaces in binary-fluid and polymer blend samples to create novel soft-composite materials. Binks productively examined the analogy between emulsification using colloids and conventional surfactants~\cite{Binks02}; the research reviewed here has taken this idea further: leading to the creation of a sponge phase analog, arrested due to interfacial colloids. The points at which the analogy breaks down, such as with the permanence and solidity of the interface, present fundamentally new directions to explore and new properties to exploit. 

\section{Acknowledgments}
Thanks to R.~Composto, E.~Herzig, T.~Horozov, R.~Sastri, J.~Thijssen, L.~Walker and K.~White for helpful comments and to A.~Morozov for checking note~\cite{Poisson}. The work in Edinburgh was also made possible by B.~Binks, M.~Cates, W.~Poon, and A.~Schofield. The microscopy facilities were provided by the Collaborative Optical Spectroscopy and Micromanipulation Centre (COSMIC). Funding in Edinburgh was provided by the EPSRC (EP/D076986/1 and EP/E502652/1) with additional support from the EU NoE SoftComp.


\begin{thebibliography}{100}
\bibitem{Poisson}
Vella and coworkers~\cite{Vella04} calculated a value for Young's modulus for a close-packed layer of colloids at a fluid-fluid interface. Following a stress the changes in dimensions are $\Delta \ell_{\|} = 2 \epsilon$; $\Delta \ell_{\bot} \approx -\frac{2}{\sqrt 3} \epsilon$. Since for hexagonal packing the unit cell is not the same size in the two orthogonal directions these are not proportional to the strains by the same factor. Instead the strains are $\gamma_{\|} = \frac{\epsilon}{R}$; $\gamma_{\bot} \approx -\frac{1}{3}\frac{\epsilon}{R}$ 
and the Poisson ratio is $\nu = \frac{|\gamma_{\bot}|}{\gamma_{\|}} \approx \frac{1}{3}$. This is a small correction to the value derived in ref.~\cite{Vella04}. The Young's modulus becomes $E \approx 7.2 \gamma / d$, a somewhat larger value which agrees less well with the experimental measurements that were presented there.
\end{thebibliography}
\end{document}